\headline{\hfill}
\magnification=\magstep1
\baselineskip 14pt

\def\Rf{\parindent=0pt\medskip\hangindent=3pc\hangafter=1}
\def\dt{{\delta t}}

\def\gtsim{\mathrel{\raise.3ex\hbox{$>$}\mkern-14mu \lower0.6ex\hbox{$\sim$}}}
\def\half{{\scriptstyle {1 \over 2}}}

\def\ltsim{\mathrel{\raise.3ex\hbox{$<$}\mkern-14mu \lower0.6ex\hbox{$\sim$}}}
\def\msun{{M_\odot}}

\def\onetwelfth{{\scriptstyle {1 \over 12}}}
\def\simgt{\gtsim}
\def\simlt{\ltsim}


\null\bigskip
\centerline{\bf Binary--single-star scattering --- VI.}
\centerline{\bf Automatic Determination of Interaction Cross Sections}
\bigskip
\centerline{Stephen L.~W.~McMillan}
\smallskip
\centerline{Department of Physics and Atmospheric Science,}
\centerline{Drexel University,}
\centerline{Philadelphia, PA 19104}
\centerline{\it steve@zonker.drexel.edu}
\medskip
\centerline{Piet Hut}
\smallskip
\centerline{Institute for Advanced Study,}
\centerline{Princeton, NJ 08540}
\centerline{\it piet@sns.ias.edu}
\bigskip

\bigskip\noindent
\centerline{\bf Abstract}
\medskip\noindent
Scattering encounters between binaries and single stars play a central
role in determining the dynamical evolution of a star cluster.  In
addition, three-body scattering can give rise to many interesting
exceptional objects: merging can produce blue stragglers; exchange can
produce binaries containing millisecond pulsars in environments quite
different from those in which the pulsars were spun up; various types
of X-ray binaries can be formed, and their activity can be either shut
off or triggered as a result of triple interactions.

To date, all published results on three-body scattering have relied on
human guidance for determining the correct parameter range for the
envelope within which to perform Monte--Carlo scattering experiments.
In this paper, we describe the first fully automatic determination of
cross sections and reaction rates for binary--single-star scattering.
Rather than relying on human inspection of pilot calculations, we have
constructed a feedback system that ensures near-optimal coverage of
parameter space while guaranteeing completeness.  We illustrate our
approach with a particular example, in which we describe the results
of a three-body encounter between three main-sequence stars of
different masses.  We provide total cross sections, as well as
branching ratios for the various different types of two-body mergers,
three-body mergers, and exchanges, both non-resonant and resonant.
The companion paper in this series, Paper VII, provides a full survey
of unequal-mass three-body scattering for hard binaries in the
point-mass limit.

\vfill\eject

{\bf 1.  Introduction}
\medskip 

The pioneer in the field of binary--single-star scattering experiments
was Hills (1975), who reported the first results in this area (see
Hills 1991 for references to subsequent papers).  In most of these
early experiments, encounters took place at zero impact parameter.
Since then, the most common way of reporting the outcome of scattering
experiments has been to quote cross sections for processes of interest
(see \S2 below).  These can then be translated into rates for use in a
variety of applications, ranging from heating and interaction rates
for Fokker-Planck and Monte-Carlo simulations, to estimates of
collision rates and branching probabilities for the production of
specific classes of object.  The first systematic direct determination
of cross sections and reaction rates for binary--single-star
encounters was made by Hut \& Bahcall (1983; Paper I in this
series---see Goodman \& Hut 1993, Paper V, for a complete set of
references to previous papers in this series).

In all these studies, the high-level organization of the
calculation---combining the results of individual calculations into
cross sections or other statistical data---has been the responsibility
of the experimenter.  Typically, for each type of total or
differential cross section, a detailed search of impact-parameter
space was performed as a pilot study before production runs were
started, mainly to determine the maximum impact parameter (i.e. the
lateral offset from a head-on collision, as measured at infinity) that
should be used.  A constant problem with this approach has been the
fact that allowing too large a value for the impact parameter can
waste a great deal of computer time on uninteresting orbits, while too
small a value will systematically underestimate some cross sections,
by missing some encounters of interest.  In this paper we present a
system for automatically determining cross sections and reaction rates
for binary--single-star scattering including, among many other
powerful features, a robust means of estimating the maximum impact
parameter to use in any given problem.

A major motivation for producing a fully automatized gravitational
scattering package for studies of the three-body scattering problem is
the observation of large numbers of primordial binaries in globular
clusters.  Until the late 1980s, it was widely believed that, while
binaries abound in open clusters (where reported binary fractions may
approach 100 percent in some cases), globular clusters were born with
few, if any, primordial binary systems.  Accordingly, it was assumed
that dynamically significant binaries had to form dynamically---by
conservative three-body interactions, or by dissipative tidal
capture---in the dense cores of evolved clusters, and that they were
few in number.  That view has since been reversed, and estimates of
the primordial binary fraction in globular clusters now range from
$\sim$3 to $\sim$30 percent, with 10--20\% regarded as fairly typical
(for a review, see Hut et al. 1992).

If binary components are drawn from the same mass distribution as
other cluster members, cluster binaries will tend to be more massive
than the average single star.  Mass segregation will then concentrate
binaries in the core, resulting in a central binary fraction
approaching 50 percent even in clusters with overall binary fractions
as low as 5 percent (McMillan \& Hut 1994).  In order-of-magnitude
terms, the mean time between significant interactions for a marginally
``hard'' binary (that is, one whose binding energy is comparable to
the mean kinetic energy ${3 \over 2} kT$ of single cluster stars) is
roughly equal to the local relaxation time, which may be only a few
million years in the densest cluster cores.  Thus we expect binaries
in star clusters to interact frequently with their environment.

Soft binaries tend to be disrupted upon encountering other cluster
members.  Hard binaries, however, are of great dynamical importance,
as their interactions tend to release energy, leaving the binaries
more tightly bound and heating the cluster (Heggie 1975).  N-body
simulations have amply demonstrated the importance of even a single
binary in controlling the overall dynamics of its parent cluster (see
Aarseth 1985, and references therein).

It now seems possible that as many as ten percent of the binaries in a
typical cluster might have orbital semi-major axes $\simlt 1$A.U., or
binding energies of $\simgt 10 kT$ (assuming a cluster 3-D velocity
dispersion of $\sim$10 km/s).  This leads to the remarkable conclusion
that the internal energy reservoir in the form of binary binding
energy may well exceed the total kinetic energy of the cluster as a
whole (in the form of center-of-mass motion of single stars and
binaries).  Numerical simulations of clusters containing substantial
numbers of primordial binaries have shown that, so long as any
binaries remain in the dynamically ``active'' energy range (1--50
$kT$, say), binaries completely control the cluster dynamics, and
continue to do so until they are all destroyed by interactions with
other binaries, or recoil out of the cluster after a triple or
four-body encounter (McMillan, Hut, \& Makino 1990, 1991; Gao et
al. 1991; Heggie \& Aarseth 1992).  For many clusters, this binary
depletion timescale may well exceed the age of universe (McMillan \&
Hut 1994).

In addition to their dynamical importance, binary interactions also
greatly increase the probability of stellar collisions, particularly
in low-density systems.  Hut \& Inagaki (1985; see also Sigurdsson \&
Phinney 1993) have shown that, during the complex orbit that
constitutes a resonant interaction, the minimum interparticle
separation is typically a few percent of the original binary
semi-major axis.  A population of 10\% 1 A.U.~binaries can increase
the stellar collision rate in a cluster by as much as two orders of
magnitude (Verbunt \& Hut 1987).  Since exchange interactions also
tend to eject the lightest of the three stars involved, we see that
binaries play a dual role: over time, they come to contain the most
massive objects in the cluster (e.g. red giants, turnoff-mass main
sequence stars, heavy white dwarfs, and neutron stars); subsequently,
they mediate interactions between these objects, possibly producing
many of the blue stragglers, millisecond pulsars, and X-ray binaries
observed in globular clusters today.

While large-scale N-body or Monte-Carlo simulations are probably
necessary to study in detail the interplay between the many different
physical processes occurring in star clusters, the very
comprehensiveness of these approaches makes it difficult to
disentangle competing physical effects.  In a sense, modelers are
placed in the position of observers, trying to infer cause and effect
from the mass of data they obtain.  Scattering experiments provide an
alternative, and much more controlled, means of investigating binary
interactions, allowing the investigator to isolate and study specific
processes in a systematic manner.  This paper is the sixth in a series
discussing various aspects of the three-body scattering problem.  Up
to now, the series has concerned itself exclusively with situations in
which the point-mass approximation is valid.  However, the software
described below has application far beyond this restricted context,
and the example we present in \S4 below includes a simple prescription
for stellar collisions.  Binary--binary interactions, which also play
a critical role in star-cluster evolution, are not considered here.
They will be the subject of a future paper.

With a few notable, and relatively recent, exceptions (Rappaport et
al. 1989; Hills 1992; Sigurdsson \& Phinney 1993; Davies 1995; Heggie
et al. 1996), most studies of binary encounters have confined
themselves to the simplest possible case of identical point masses.
Expanding our attention to cover unequal masses and finite-radius
effects requires us to change at a very fundamental level the way
scattering calculations are performed.  In the simple case, it is
possible to publish ``atlases'' of encounter outcomes covering most of
the parameter space of interest (e.g.~Hut 1984). However, in the
general case, there are simply too many parameter combinations for
such atlases to be feasible, or even desirable.  Instead, specific
questions generally require individual calculations.  Rather than
having a standard format for reporting cross sections, it is much more
useful to have a standard, reliable means of obtaining new results as
needed.

In this paper we describe a software system for performing these
calculations in a robust and efficient manner, automating the many
error-prone steps needed to perform a series of scattering experiments
and arrive at a useful answer in the form of a set of cross sections
or distribution of outcomes.  Two ingredients have made these new
developments possible.  One is the increase in available computer
speed: Computational capabilities increased by more than two orders of
magnitude between the earliest scattering experiments (Hills 1975) to
those reported in Paper I, and speeds have increased by a similar
factor between then and now.  The second ingredient is the wide
availability of high-level computer languages that enable and invite a
flexible and modular approach to programming, making automation far
easier.

The present paper is arranged as follows.  Section 2 discusses the
general procedures for determining gravitational scattering cross
sections.  Section 3 describes the STARLAB software environment, with
particular reference to the automatization techniques we employ.
Section 4 presents a simple example illustrating the use of the system
in practice.  A more substantial application of the package, in the
determination of exchange cross sections in the general problem of
binary--single-star scattering with unequal masses, is the subject of
the next paper in this series (Heggie, Hut \& McMillan 1996; Paper
VII).  Another recent application concerns the formation of triple
systems in binary-binary encounters, in the limit where the tightest
of the two binaries can be approximated as a single point mass (Rasio,
McMillan \& Hut 1995).  Finally, in Section 5, we summarize and
conclude.

\bigskip
\noindent
{\bf 2.  Computation of Cross Sections}
\medskip

For a detailed description of how to perform and analyze scattering
experiments, see Paper I.  We limit ourselves here to a brief summary
of the parameters and conventions that have been used in the series so
far.  For a brief history of gravitational scattering experiments, see
Hut (1992; Paper III).  For more recent contributions, see Sigurdsson
\& Phinney (1995) and Davies (1995).

In the point-mass approximation, if the system center of mass is taken
to be at rest at the origin, 15 parameters are required to specify a
scattering encounter.  The target binary's internal parameters are the
semi-major axis $a$, eccentricity $e$, component masses $m_1$ and
$m_2$, and initial mean anomaly $M$.  Three more angles specify the
binary's orbital plane and the orientation of the major axis within
that plane.  Without loss of generality, we can take the binary to lie
in the $x$--$y$ plane, with its major axis along the $x$-axis.  The
remaining 7 parameters specify the mass $m_3$ of the incomer and its
trajectory relative to the binary center of mass.  Again, three angles
determine the orientation of the incomer's orbit, while its initial
phase, energy and angular momentum are conveniently set by choosing
the separation $R_3$, relative velocity at infinity $v_\infty$ and
impact parameter $\rho$.

A scattering experiment entails integrating a large number of
individual three-body encounters, holding some parameters fixed and
choosing others randomly from specified distributions.  In a typical
case, the binary parameters $a$, $m_1$ and $m_2$ are held fixed, $M$
is chosen uniformly on $[0, 2\pi)$, while the eccentricity $e$ is
either held fixed or chosen randomly from a thermal distribution $f(e)
= 2e$.  For the outer orbit, $m_3$ and $v_\infty$ are held fixed, the
orbital orientation is chosen randomly, $\rho$ is chosen uniformly in
$\rho^2$ between $\rho = 0$ and $\rho = \rho_{max}$ (a parameter whose
value will be discussed in a moment), and $R_3$ is chosen to keep the
initial tidal perturbation below some tolerance $\gamma_{min} \ll 1$:
$R_3\sim a\gamma_{min}^{-1/3} [1 + m_3/(m_1+m_2)]^{1/3}$ (the mass
factor ensuring that $R_3$ will be substantially larger for, say, an
incoming $10 \msun$ black hole than for a $0.5 \msun$ dwarf).  In
practice, it is convenient to adopt units such that $G$, $a$, and $m_1
+ m_2$ are all equal to unity, so the binary period is $2\pi$.

If the relative velocity at infinity is less than the critical
velocity $v_c$, defined by
$$
    v_c^2 \equiv {G m_1 m_2 (m_1 + m_2 + m_3) \over (m_1 + m_2) m_3 a}\,,
	\eqno{(1)}
$$
the total energy is negative and the scattering (in the absence of
collisions) will eventually result in a binary and an unbound single
star.  The calculation is terminated when the perturbation on the
binary again falls below $\gamma_{min}$.  If $v_\infty \ge v_c$,
disruption of the initial binary into three unbound stars is also
energetically possible.  In that case, the calculation stops when all
three stars are mutually unbound and none is a significant perturber
on the relative motion of the other two.

Once the integration is over, the encounter is classified as a
resonance or a non-resonance, depending on the number of minima
$N_{min}$ in the quantity $s^2 = r_{12}^2 + r_{23}^2 + r_{31}^2$,
where $r_{ij}$ is the instantaneous separation between particles $i$
and $j$.  For very early and very late times, $s^2$ takes on
arbitrarily large values, since at least one of the three stars will
be far removed from the other two.  At some point during the
scattering experiment, $s^2$ will take on a minimum value.  In a {\it
non-resonant} interaction, $s^2$ shows only one local minimum; {\it
resonant} interactions are characterized by more oscillatory behavior,
with $s^2$ taking on several local minima.  In practice, this simple
recipe can be too sensitive to the behavior of $s^2$; we have found it
safer to accept two successive local minima as distinct only if, at
the intervening maximum, the value of $s^2$ is at least twice that at
both minima.  Without this additional condition, oscillations in $s^2$
can lead to significant numbers of non-resonant encounters being
erroneously classified as resonances.

Resonances are further subdivided into ``democratic'' and
``hierarchical'' categories, as discussed in Papers III and VII.  In
some circumstances, we have found it convenient to refine the
classification still further based on the value of $N_{min}$.  The
final state of the system is classified as a preservation, if the
initial incomer escapes; an exchange, if one of the binary components
escapes; or an ionization, if the binary is destroyed.  If nonzero
stellar radii are included, physical collisions between stars become
possible and additional final states must be defined.  A complete list
of intermediate and final states is presented in Table 1.

Cross sections are calculated in the usual way (cf.~Paper I).
Specifically, in the case of uniform sampling with $N$ trials between
impact parameters $\rho = 0$ and $\rho = \rho_{max}$, the cross
section for events of type X is
$$
    \sigma_X = \pi \rho_{max}^2 {N_X \over N}  \,,
	\eqno{(2)}
$$
where $N_X$ is the number of times outcome X occurs.  Differential
cross sections (e.g.~$d\sigma/dE$, where $E$ is the final binary
energy) are determined in an analogous way.  The standard error in
$\sigma_X$ is given by
$$
    \delta\sigma_X = {\sigma_X \over \max(1, \sqrt{N_X})}  \,.
	\eqno{(3)}
$$
The sampling continues until $\delta\sigma_X$ for the process of
interest falls within some specified tolerance.

The procedure outlined above is straightforward, if tedious, but it
suffers from a number of practical problems.  It is fairly simple to
write a program to perform the steps just listed, but it is also very
easy for sampling and other systematic errors to creep into the
calculation.  A major reason for this is the fact that these
computations are typically performed on several machines in parallel
over a period of hours or days, each machine covering come portion of
parameter space independently of the others.  The programmer then
collates and analyzes the results after the fact.  This procedure is
particularly prone to errors when it becomes necessary to extend the
sampling, and great care must be taken to ensure that uniformity is
maintained.  However, probably the most common source of error is an
incorrect choice of $\rho_{max}$, whose value depends both on the
particular cross section(s) of interest and on the overall accuracy
desired.  Errors in the choice of $\rho_{max}$ can be very difficult
to detect.

Sadly, successful negotiation of all these hurdles in one calculation
is no guarantee of success in the next, when parameters are changed
and termination criteria may be subtly altered.  If one wishes to
determine many cross sections with reasonable confidence that the
integrations, sampling, and statistics are all being properly handled,
it is clear that an automated procedure of some sort is highly
desirable.  In addition, once in place, such a system should allow the
user to ask ``high-level'' questions (i.e. about astrophysics) without
having to be unduly concerned with the minutiae of the calculation.
We now describe the basic elements of just such a system.

\bigskip
\noindent
{\bf 3.  Scattering Experiments in STARLAB}
\medskip
\noindent
{\bf 3.1  STARLAB}
\smallskip

The scattering package described here operates within the STARLAB
software environment.  STARLAB is a collection of modular software
tools for simulating the evolution of dense stellar systems and
analyzing the resultant data.  It consists of a library of loosely
coupled programs, sharing a common data structure, which can be
combined in arbitrarily complex ways to study the dynamics of star
clusters and galactic nuclei.  Individual modules may be linked in the
``traditional'' way, as function calls to C++ (the language in which
most of the package is written), C, or FORTRAN routines, or at a much
higher level---as individual programs connected by UNIX pipes.  The
former linkage is more efficient, and allows finer control of the
package's capabilities; however, the latter provides a quick and
compact way of running small test simulations.  The combination
affords great flexibility to STARLAB, allowing it to be used by both
the novice and the expert programmer with equal ease.

In the present context, a single scattering experiment might be
performed as follows:

\medskip\centerline{\tt
mkscat -n 3 -v 0.1 | low\_n\_evolve -a 0.02 -D 0.1 | xstarplot -o | analyze
}\medskip\noindent
Here {\tt mkscat -n 3} creates initial conditions for a three-body
scattering calculation, in this case with a velocity at infinity of
$0.1 v_c$ (all parameters, such as $\gamma_{min}$, $\rho$, and so on,
may be specified in a similar manner).  The resulting data are passed
to the next module, {\tt low\_n\_evolve}, which integrates the
equations of motion with an accuracy parameter of 0.02 until the
encounter is over.  The next link is a display program that allows the
user to view the encounter in real time, updating the display every
0.1 time units (the output frequency specified with the ``{\tt-D}''
option in the invocation of {\tt low\_n\_evolve}).  Finally, the
module {\tt analyze} classifies the encounter into one of the
categories described in Table 1, and produces some diagnostic output.
While the vertical bars between commands in the above example
represent UNIX pipes, they could equally well stand for C++ function
calls in a single compiled program---the design of the software makes
the distinction largely irrelevant.

The full STARLAB package contains three groups of programs for use in
studies of stellar dynamics, hydrodynamics, and stellar evolution.
The separation between these three groups is not rigorous, since some
programs act as bridges between these different parts.  For example, a
simulation of stars with finite radii and finite lifetimes can be
driven by an integrator that computes the orbits of the bodies in the
point-mass limit.  Whenever two of these stars come close enough
together to exhibit non-point-mass interactions, including tidal
distortions and physical collisions, the stellar dynamics module
notifies the hydrodynamics module to do the required additional
computations.  Similarly, when time progresses beyond the scheduled
main-sequence lifetime of a star, a stellar evolution module can keep
track of the increasing radius of the star as it climbs the giant
branch.

A persistent problem in N-body (and many other) simulations is the
question of how to handle data outside the scope of a particular
program or module.  For example, while it is generally not necessary
to consider stellar evolution when performing scattering experiments,
collisions are fairly commonplace, and this requires some treatment of
hydrodynamic interactions.  The modularity of the system makes it
highly desirable that the hydrodynamics programs not know or rely upon
the details of the particular dynamical integrator being used, and
vice versa.  Furthermore, collisions clearly affect the subsequent
evolution of the stars involved, but stellar-evolutionary programs
need know nothing of the dynamical events preceding the formation of
the stars being studied.  (As a practical matter, it is also quite
common that two or more different investigators, each unfamiliar with
the specifics of the others' areas, are involved, making it all the
more important that extraneous details be suppressed.)

A unique aspect of STARLAB is the fact that its internal data
structure and external data representation are specifically designed
to prevent ``unknown'' information from being lost or corrupted as it
moves around the system.  Each individual body in the system is
represented as a node in a tree constructed to reflect the existence
of closely interacting subsystems and any internal stellar structure.
Information unknown to a particular module is simply stored as
character data attached to a node, to be reproduced at the time of
output in the correct format and location, preserving the
correspondence of the data with the body in question.  This allows the
use of an arbitrary combination of pipes, while guaranteeing that data
and comments are preserved.  In addition, a list of all commands
used to create the data, together with the time at which the commands
were issued, is stored as part of the data stream, to minimize any
uncertainty about the exact procedures used.

For example, the following command string performs a simple N-body
calculation:
\medskip\centerline{\tt
mkplummer -n 100 | evolve -t 10 | mark\_core | HR\_plot | evolve
}\medskip\noindent
A 100-body Plummer model is generated and evolved forward in time for
10 time units.  The module {\tt evolve} integrates the dynamical
equations of motion, at the same time following the ways in which
individual stars age and interact hydrodynamically (these functions
being performed by separate STARLAB modules compiled into the
program).  The program {\tt mark\_core} computes the location and size
of the core of the star cluster, printing out some diagnostic
messages.  The fourth module, {\tt HR\_plot}, plots a
Hertzsprung-Russell diagram of the star cluster (perhaps using special
symbols for the core stars), before returning control to the module
that evolves the whole system.  For this to work, {\tt mark\_core}
must preserve the stellar-evolution information, even though it only
``knows'' about the dynamical part of the data.  Similarly, {\tt
HR\_plot} preserves dynamical data, even though it never references
(explicitly or implicitly) any dynamical quantity.  The bookkeeping
required to do all this is performed by low-level STARLAB I/O
routines, and is completely transparent to the user.

\medskip
\noindent
{\bf 3.2  Orbit Integration}
\smallskip

Our orbit integration scheme is that of the fourth-order
variable--time-step Hermite integrator described by Makino \& Aarseth
(1992).  This scheme is the natural generalization of the familiar
leapfrog scheme to higher order (see Hut, Makino \& McMillan 1995).
The algorithm to advance the position $x$ and velocity $v$ from time 0
to time 1, with time step $\dt$, may be conveniently expressed as
follows:
$$
x_1 = x_0 + \half(v_1 + v_0)\dt - \onetwelfth(a_1-a_0)(\dt)^2 \,,
	\eqno{(4)}
$$
$$
v_1 = v_0 + \half(a_1 + a_0)\dt - \onetwelfth(j_1-j_0)(\dt)^2 \,,
	\eqno{(5)}
$$
\smallskip\noindent
where $a$ is acceleration and $j = da/dt$.  The above, formulation,
which appears to be implicit as stated, is actually implemented as a
predictor-corrector scheme in the usual way (Makino
\& Aarseth 1992).

A drawback of this and similar schemes when applied to scattering
problems, where lightly perturbed binaries must be followed for many
thousands of orbits, is the fact that it makes a small but systematic
error in the integration of periodic orbits, and these errors can
become significant over the time required for the incoming star to
reach the target binary.  To control these errors, the STARLAB
scattering integrators make use of a novel technique developed by Hut,
Makino \& McMillan (1995) which guarantees time symmetry in the
integration and hence enforces exact energy and momentum conservation
in periodic orbits.  This results in spectacular improvements in
long-term stability of the integration scheme in all circumstances,
even in the case of long-lived resonances.  Briefly, the actual time
step $\dt$ is forced to be a symmetric function of the initial (``0'')
and final (``1'') states in each step, simply by iterating on the
timestep until the condition
$$
	\dt = \half [\Delta t(x_0, v_0), \Delta t(x_1, v_1)]
	\eqno{(6)}
$$
is satisfied, where $\Delta t(x, v)$ is the ``natural'' time step
corresponding to position $x$ and velocity $v$.  In practice, we find
that one or two iterations are sufficient, and that the gain in
accuracy and stability greatly outweigh the extra computational effort
required.

In addition, we make use of analytical approximations for the orbits
of the three stars in cases where one of the stars makes a distant
excursion.  If the strength of the tidal perturbation of the third
stars drops below a certain threshold, an attempt is made to continue
its trajectory analytically along a two-body Kepler orbit.  To this
end, the remaining two stars are replaced by a single particle
positioned at their center of mass, with mass equal to the total mass
of the pair.  There are three possible outcomes for each attempt:

{\narrower

\item{(1)}{If the Kepler orbit thus found is clearly hyperbolic, the
scattering experiment is considered to be over.}

\item{(2)}{If the orbit is clearly elliptic, the third star is placed
on the incoming portion of the elliptic orbit, at roughly the same
distance from the binary as at the onset of the analytic
approximation.  The exact position is determined by requiring that the
total duration of the analytic approximation phase be an integral
number of inner orbital periods.  As a result, the error produced by
neglecting the tidal interaction at the onset of the analytic phase is
nearly canceled by the similar error made at the end of this phase.
This has the additional practical benefit that the relative position
of the inner two stars does not have to be recomputed.}

\item{(3)}{If the Kepler orbit is nearly parabolic, no analytic
approximation phase is initiated.  Rather, numerical integration is
continued for a fixed amount of additional time, after which another
attempt is made to fit a Kepler orbit.  This procedure forms an
important safeguard against classifying an experiment as being
finished on the basis of either small numerical errors or small
physical effects stemming from the neglect of tidal perturbations.}

}

This implementation speeds up the calculations significantly (for an
earlier description and analysis of the time gain, see Papers I and
III).  However, occasional rare two-body encounters at extremely close
distances can sometimes lead to errors exceeding our tolerance
(typically a change in total energy larger than a fraction $10^{-4}$
of the binding energy of the original binary).  To remedy these
situations, a second analytic approximation is used in which both the
inner and outer orbits are replaced by Kepler orbit segments whenever
any portion of the inner orbit can safely be regarded as unperturbed.
Note that the trigger criterion here differs from that used in the
previous case of long excursions, where the distance to the third body
greatly exceeds the semi-major axis of the inner binary.  While the
same perturbation threshhold is used in either case, the analytic
phase here covers only a short time interval centered on the
periastron of the inner orbit.

\medskip
\noindent
{\bf 3.3  Automatic Scattering Experiments}
\smallskip

The automated scattering software in STARLAB is constructed in several
layers, each largely independent of the others.  While the system is
designed to function with only high-level user input, each layer
remains accessible if necessary, and arbitrarily complex diagnostic
data---from graphical output to user-defined functions---at each level
can be obtained as desired, without the need for rewriting and
recompiling existing sections of code.  The layers are as follows:

\smallskip\noindent 1.  Apart from the basic I/O and data-handling
routines that form the foundation of the entire STARLAB package, the
lowest-level of the scattering subsystem consists of an orbit
integration engine, using the integration scheme and analytical
approximations described in \S 3.2.

\noindent 2. Above the integration scheme lies the basic ``workhorse''
integrator, {\tt low\_n\_evolve}, designed to advance an arbitrary
N-body system through a specified time interval $T$, with variable
(time-symmetrized) time steps and numerous built-in diagnostic and
consistency checks.

\noindent 3. On top of these lowest levels lie several
scattering-specific layers.  These include: (1) routines to create an
initial scattering state, holding some parameters fixed while choosing
others randomly, as discussed in \S 2; (2) checks to determine whether
a given scattering experiment has reached its final outgoing state;
(3) optimization features, such as analytical integration of inner and
outer orbits of hierarchical triple systems in which the outer orbital
period vastly exceeds the inner orbital period, or assuming
unperturbed two-body motion near the pericenter of a close encounter;
(4) diagnostic functions to store information describing the buildup
of energy errors; (5) various bookkeeping functions to chart the
overall character of the orbits (e.g.~democratic versus hierarchical
resonance states); (6) checks for overlap of stellar radii
(implemented as user-defined function calls from the integrator), in
which case merging routines are invoked to replace colliding stars
with a single merger product.

With the optimizations (3) above, we have found it unnecessary to
employ elaborate regularization techniques such as
Kustaanheimo--Stiefel regularization (Kustaanheimo \& Stiefel 1965;
Aarseth 1985), whose complex formulation greatly complicates the
integration (but see Funato et al. 1995).  All checks (termination,
bookkeeping, etc.) are applied after each invocation of {\tt
low\_n\_evolve}, with the exception of (6), which is applied at the
end of every time step within {\tt low\_n\_evolve} itself.  Typically,
the time interval $T$ is taken to be 10--20 dynamical time units, or
about 2--3 initial binary orbits.

\noindent 4. Atop these layers lies the first user-accessible module
(in normal use): {\tt scatter3}, a function to initialize, integrate,
and classify a single three-body scattering, with a large number of
built-in options.  The masses and radii of the stars can be specified,
as well as the orbital parameters of the binary, the impact parameter
of the encounter, and the relative velocity, asymptotically far before
the encounter.  The initial distance from which the integration starts
is determined automatically, with a default $\gamma_{min}$ of
$10^{-6}$.  In addition, an overall accuracy parameter controls the
cost/performance ratio.  In practice, typical relative energy errors
can be easily kept as small as $10^{-10}$; our production runs usually
aim at median errors of order $10^{-6}$, allowing a speed-up of a
factor of ten in computer time with respect to the most accurate
integrations we can achieve.

\noindent 5. The next layer is the module {\tt sigma3}, which
contains all the management software needed to conduct a complete
series of scattering experiments.  Depending on the type of total or
differential cross section requested, the user can choose an
appropriate command to activate a ``beam'' of single stars aimed at
the ``target plate'' of binary stars.  However, rather than relying on
human inspection of pilot calculations, the STARLAB package monitors
its own progress as the computation proceeds, and adjusts itself to
ensure proper coverage of parameter space.

We start by performing $n$ scatterings (where the ``trial density''
$n$ is a user-specified parameter), uniformly distributed in impact
parameter over the range $0\le\rho<\rho_0 = 2a[1+G(m_1+m_2+m_3) /
av_\infty^2]^{1/2}$.  The value of $\rho_0$ simply corresponds to a
periastron separation of $2a$.  The impact parameter range is then
systematically expanded, covering successive annuli of outer radii
$\rho_i = 2^{i/2}\rho_0$ with $n$ trials each, until no interesting
interactions take place in the outermost zone ($i=i_{max}$) sampled.
Typically, an ``interesting'' interaction is one in which the binary
is significantly perturbed in energy or eccentricity, although the
precise definition may be modified as desired.  This procedure
produces rapid convergence toward accurate cross sections, with a
minimum of wasted effort.  We note that this simple prescription will
probably fail for cross sections with significant off-axis peaks, but
the method could be easily generalized if necessary.

With non-uniform sampling, the earlier expressions (Eq.~(2) and
(3)) for the cross sections and errors become:
$$
    \sigma_X		=  {\pi \over n} \sum_{i=0}^{i_{max}}
						(\rho_i^2 - \rho_{i-1}^2)
							N_{Xi} \,,
	\eqno{(7)}
$$
and
$$
    (\delta\sigma_X)^2	=  {\pi^2 \over n^2} \sum_{i=0}^{i_{max}}
				(\rho_i^2 - \rho_{i-1}^2)^2 N_{Xi} \,,
	\eqno{(8)}
$$
where $N_{Xi}$ is the number of times outcome X occurs in zone $i$,
and $\rho_{-1} = 0$.

The standard output from {\tt sigma3} is a table of total cross
sections and errors for all possible combinations of ``intermediate''
and ``final'' states (see Table 1).  In typical use, we begin with a
low-density (low-$n$) calculation, producing a preliminary report,
with estimates of all relevant cross sections plus their corresponding
error bars, after a few minutes.  Thereafter, the density is increased
by factors of four until the desired value is reached, so each
subsequent report appears after a four times longer interval.  Because
of the Monte-Carlo nature of the impact-parameter sampling, each
successive report carries error bars that are half the size of the
corresponding ones in the previous report.  Thus reasonable estimates
can often be obtained in ten or fifteen minutes, with more accurate
results following in an hour (on a fast workstation).

The design of the scattering package also allows it to be used as a
basis for more sophisticated statistical studies.  The system can
invoke user-written functions at key points during and after each
scattering event, providing the user with a correctly sampled
environment in which other, more elaborate, calculations may be
performed.  The user's routines receive all pertinent data on each
scattering, along with the appropriate statistical weight associated
with the event.  In this way, any desired information may be collated
and analyzed without concern for (or even knowledge of) the mechanics
of setting up the individual interactions, or of the details of the
sampling.

\noindent 6. The present highest-level layer, on top of the cross
section manager, is the Maxwellian rate estimator {\tt rate3}.  After
specification of the stellar characteristics, the binary orbit, and
the velocity dispersion of the single stars, this estimator
automatically computes cross sections for processes of interest at
different points under the Maxwellian velocity distribution curve,
multiplying the results by the Maxwellian weight factor, and adding
those to obtain reaction rates.  As before, the longer one is willing
to wait, the more accurate the rates become (through an automatic
increase both in the number of velocity points, and in the accuracy
of the cross sections determined at each).

Still more layers can be added with little additional investment in
time.  With the complexity of orbit integration and scattering
management hidden in the various modules, it is relatively
straightforward to implement new levels.  For example, one might wish
to address the inverse scattering problem: given a final system, what
is the relative probability that such a system originated from
different initials conditions?  This could be handled as follows:
after specifying the velocity dispersion and other stellar parameters,
an automatic tabulation of Maxwellian reaction rates could be
performed, while filtering the results to allow only those scattering
experiments to be counted that led to the desired range of final
parameters.  This is essentially the approach followed by Rasio et
al.~(1995).

\noindent 7.  Parallel (PVM) implementations of {\tt sigma3} and {\tt
rate3} are currently under development.

\bigskip\noindent
{\bf 4.  A Simple Application}
\medskip

The most basic application of the STARLAB scattering package is the
generation of total cross sections for specific processes of
astrophysical interest.  Here we present cross sections for physical
collisions and non-colliding exchanges and resonances during
encounters between main-sequence binaries and incoming stars, for
parameters typical of the core of an evolved globular cluster.

The binary components are taken to have masses $m_1 = 0.8$ and $m_2 =
0.4 M_\odot$, and radii $R_1 = 0.8$ and $R_2 = 0.4 R_\odot$,
respectively.  The initial binary orbital eccentricity is randomly
chosen from a thermal distribution, with the proviso that the
separation at periastron is at least twice the sum of the stellar
radii, so that immediate collisions are avoided.  The incomer is taken
to be an intermediate-mass main-sequence star, of mass $m_3 = 0.6
M_\odot$, radius $R_3 = 0.6 R_\odot$, and velocity at infinity
$10\,{\rm km\,s}^{-1}$.  The results reported here are intended mainly
to illustrate the capabilities of the software; they may be compared
with similar calculations presented elsewhere in the literature, most
recently by Davies (1995).

We have chosen to consider this particular case because the outcome of
a collision between two main-sequence stars is fairly well known: if
the two stars approach within roughly the sum of their radii, they
merge to form a single object of approximately double the original
radius, with negligible mass loss (see, e.g., Benz \& Hills 1987;
Lombardi et al. 1995).  In order to determine whether or not a second
merger occurs (if the first occurs with the third ``spectator'' star
bound to the center of mass of the colliding pair), we assume the
simple mass--radius relation $R\propto m$.  The arguably more
interesting case of encounters involving compact binary components or
incomers is not considered here, because the lifetimes, properties,
and appearance of collision products are not sufficiently (if at all)
well known to be easily distilled into a simple illustrative example.

The possible results of an encounter then are: (1) a non-colliding
(``clean'') exchange, preservation, or ionization; (2) a two-body
merger, leaving a $1.0$, $1.2$, or $1.4 M_\odot$ blue straggler, which
may itself be part of a stable binary system; (3) a triple merger,
forming a $1.8 M_\odot$ blue straggler.  Fig.~1 shows the cross
sections for these processes (excluding preservations, whose cross
section is obviously infinite), for binary semi-major axes ranging
from $0.02$ to $100$ A.U.

In this figure, each set of normalized cross sections, along with
error estimates, is generated by a single invocation of {\tt sigma3}
of the form:
\medskip\centerline{\tt
    sigma3 -m 0.3333 -M 0.5 -v $v$  -r1 $r_1$  -r2 $r_2$ -r3 $r_3$  -d 2500.
}
\medskip\noindent
The binary mass $m_1+m_2$ and semi-major axis $a$ define the mass and
length units of the calculation, the ``{\tt-m}'' and ``{\tt-M}''
switches specify the binary secondary and incomer masses,
respectively, $v = 10\,{\rm km\,s}^{-1}$ $/\ v_c(0.8 M_\odot,$ $0.4
M_\odot,$ $0.6 M_\odot,\ a)$ (where $v_c$ is defined in Eq.~(1)), and
$r_i = R_i / a$.  The ``{\tt-d}'' switch sets the trial density.  With
this choice of parameters, generation of each set of points took
$\sim\,$3--4 hours to generate on an HP-735 workstation.  The
displayed data are precisely as output by the program; here and below,
error bars are shown only where they exceed the size of the symbols
used.

The suppression of clean exchanges (whose scaled cross section should
be roughly constant in the point-mass limit) by collisions during
close encounters is clearly evident.  Ionization occurs only for
semi-major axes such that $v_c(0.8 M_\odot,$ $0.4 M_\odot,$ $0.6
M_\odot,\ a) < 10 {\rm km\,s}^{-1}$, or $a > 7.1$ A.U., as indicated
by grey dashed lines on all figures.  The break in the slope of the
``2-merger'' cross section at $a\sim10$A.U. is a direct result of our
particular choice of initial binary eccentricities, which always
permitted ``almost colliding'' systems, independent of the value of
$a$.  Collisions with $a>10$ A.U.~are mainly ``induced mergers,'' in
which the components of a very eccentric binary are perturbed onto a
collision course by the passage of the third star.  Had we begun our
simulations with circular binaries, the overall induced merger rate
would have been substantially reduced (even for $a < 10$ A.U.), and no
break in the cross section would have been evident.

For the adopted set of initial parameters, collisions dominate over
``clean'' encounters for $a\simlt0.2$A.U., corresponding to initial
binary periods of $\simlt 30$ days.  When the factor of $a$ implicit
in the $a^2 v_c^2$ scaling is taken into account, the 2-body collision
cross section is found to be roughly independent of $a$, at $\sim$20
A.U.$^2$ for 0.2 A.U.~$\simlt a\simlt 10$ A.U.  Within this range, the
merger (i.e.~``blue straggler formation'') rate {\it per binary} in a
cluster core of density $10^4 n_4$ stars pc$^{-3}$ is
$$
	R_2 \sim 0.05 n_4 {\rm Gyr}^{-1}\,.
$$
The constancy of the 2-body merger cross section is easily understood
as the combination of two factors: the overall binary interaction
cross section scales as $a$ because of gravitational focusing, while
the probability of a collision during the course of an interaction
scales as $\langle R_\ast\rangle/a\propto 1/a$.

More detailed inspection of the output from {\tt sigma3} reveals a
wealth of additional data.  Fig.~2(a) divides the non-colliding
exchanges into ``exchange 1'' events, in which star 1 (the $0.8
M_\odot$ component) escapes, and ``exchange 2'' events, where star 2
($0.4 M_\odot$) does.  The low-$a$ results are consistent with
asymptotic estimates of the exchange cross sections (see Paper VII) in
the limit $v/v_c\rightarrow0$.  (Notice, however, the crossover in the
branching ratios for high-speed encounters.)  Fig.~2(b) further
subdivides the data into non-resonant and resonant encounters.  The
fraction of resonant exchanges decreases at small $a$ because of
mergers.

For colliding encounters, Fig.~3 shows branching ratios for all
possible two-body mergers, along with the triple merger cross section,
all normalized to the total two-body merger cross section.  Triple
collisions account for only a negligible fraction ($\simlt5$\%) of the
total, except for $a = 0.02$ A.U., where they represent about 15\% of
all mergers.  The dominance of ``1+2'' mergers is due in part to
induced mergers.  The other merger events---2+3 and 1+3---occur at a
significant rate only in democratic resonances.  However, we find that
1+2 mergers also tend to be favored in resonant encounters, at least
for this particular choice of masses.

Finally, Fig.~4 shows the fraction of two-body mergers resulting in an
unbound final system (i.e. isolated blue stragglers).  Only induced
mergers are likely to be unbound; resonant mergers tend to lead to a
bound final system.

\bigskip\noindent
{\bf 5.  Summary and Conclusion}
\medskip

We have described here only one simple application of the STARLAB
scattering software. For use of the {\tt sigma3} and {\tt rate3}
programs to provide correctly sampled environments for calculations of
differential cross sections and other scattering-related quantities,
see Rasio, McMillan \& Hut (1995: the formation of the millisecond
pulsar triple system B1620-26 in M4) and McMillan (1996: dependence of
binary heating on incomer mass and impact parameter).  Finally,
Heggie, McMillan \& Hut (1996; Paper VII) have used {\tt sigma3} in
fully automated mode as the basis for an extensive calculation of
exchange cross sections, using numerical results to calibrate analytic
asymptotic expressions, to yield a fitting formula for the exchange
cross section for hard binaries, valid to about 20\% for arbitrary
masses.

The design of STARLAB facilitates inclusion of more detailed physical
processes into our models, and this represents one obvious direction
of future development of the system.  In addition, binary-binary and
general $N$-body scattering packages are nearing completion.  While
more complex, their conceptual framework is similar to that described
above for the 3-body case.  The entire package, is freely available by
anonymous FTP from ftp.sns.ias.edu/pub/starlab.  It has been
successfully installed on UNIX systems running SunOS 4, Solaris 2,
HP-UX, Linux, and Dec OSF, using both native C++ compilers and the GNU
g++ compiler.  Real-time demonstrations of the software are available
through the URL http://www.sns.ias.edu/$\scriptstyle\sim$starlab.

\bigskip\noindent This work was supported in part by NASA grant
NAGW-2559 and NSF grant AST-9308005.

\par\vfill\eject

\bigskip\noindent
{\bf References}
\medskip
{\parindent=0pt
\Rf Aarseth, S.J. 1985, in Multiple Time Scales, ed. J.U. Brackbill
	and B.I. Cohen (New York: Academic), p. 377
\Rf Benz, W., \& Hills, J.G. 1987, ApJ, 323, 614
\Rf Davies, M.B. 1995, MNRAS, 276, 887
\Rf Funato, Y., Hut, P., McMillan, S.L.W., \& Makino, J. 1995, submitted to AJ
\Rf Gao, Goodman, J.G., Cohn, H.N., \& Murphy, B. 1991, ApJ, 370, 567
\Rf Goodman, J.G. \& Hut, P., 1993, ApJ, 403, 271 [Paper V]
\Rf Heggie, D.C., 1975, MNRAS, 173, 729
\Rf Heggie, D.C. \& Aarseth S. J. 1992, MNRAS, 257, 513
\Rf Heggie, D.C., Hut, P. \& McMillan, S., 1996, ApJ, submitted [Paper VII]
\Rf Hills, J.G., 1975, AJ 80, 809
\Rf Hills, J.G. 1991, AJ, 102, 704
\Rf Hills, J.G. 1992, AJ, 103, 1955
\Rf Hut, P. 1984, ApJS, 55, 301
\Rf Hut, P. 1992, ApJ, 403, 256 [Paper III]
\Rf Hut, P. \& Bahcall, J. N. 1983, ApJ, 268, 319 [Paper I]
\Rf Hut, P., \& Inagaki, S. 1985, ApJ, 298, 502
\Rf Hut, P., McMillan, S.L.W., Goodman, J.G., Mateo, M., Phinney, E.S.,
	Pryor, C., Richer, H.B., Verbunt, F., \& Weinberg, M. 1992,
	PASP, 105, 981
\Rf Hut, P., Makino, J., \& McMillan, S.L.W. 1995, ApJL, 443, L93
\Rf Kustaanheimo, P. \& Stiefel, E.L. 1965, J. Reine Angew. Math. 218, 204
\Rf Lombardi, J.C., Rasio, F.A., \& Shapiro, S.L. 1995, CRSR preprint 1102
\Rf Makino, J., \& Aarseth, S.J. 1992, PASJ, 44, 141
\Rf McMillan, S.L.W., Hut, P., \& Makino, J. 1990, ApJ, 362, 522
\Rf McMillan, S.L.W., Hut, P., \& Makino, J. 1991, ApJ, 372, 111
\Rf McMillan, S.L.W., \& Hut, P. 1994, ApJ, 427, 793
\Rf McMillan, S.L.W. 1996, in preparation
\Rf Rappaport, S., Putney, A., \& Verbunt, F. 1989, ApJ, 345, 210
\Rf Rasio, F.A. 1994, ApJL, 427, L107
\Rf Rasio, F.A., McMillan, S.L.W. \& Hut, P. 1995, ApJL, 438, L33
\Rf Sigurdsson, S., \& Phinney, E.S. 1993, ApJ, 415, 631
\Rf Sigurdsson, S., \& Phinney, E.S. 1995, ApjS, 99, 609
\Rf Verbunt, F. \& Hut, P., 1987, in The Origin and Evolution of
	Neutron Stars, I.A.U. Symp. 125, eds. D. Helfand and
	J. Huang, (Dordrecht: Reidel), p. 187 }

\par\vfill\eject

\null\bigskip\noindent
{\bf Table 1: Three-body Scattering Outcomes}
\bigskip

\vbox{\halign to \hsize{
#\hfil&#\hfil&~~~~#\hfil\cr
\noalign{\hrule\smallskip}
Intermediate States$^\dagger$&Final States$^\dagger$&Description\cr
\noalign{\smallskip\hrule\smallskip}
\tt non\_resonance&&Single minimum in the total interparticle\cr
		  &&separation ($N_{min} = 1$)\cr
\tt hierarchical\_resonance&&Multiple minima ($N_{min} > 1$),\cr
			   &&repeated flybys by the same star\cr
\tt democratic\_resonance&&Multiple minima ($N_{min} > 1$),\cr
			 &&no preferred ``third star'' during excursions\cr
\noalign{\medskip}
&\tt preservation&Original pair remains bound\cr
&\tt exchange\_1&Binary component 1 ejected\cr
&\tt exchange\_2&Binary component 2 ejected\cr
&\tt ionization&Binary destroyed\cr
&\tt merger\_binary\_1&Stars 2 and 3 merge, 1 remains bound\cr
&\tt merger\_binary\_2&Stars 1 and 3 merge, 2 remains bound\cr
&\tt merger\_binary\_3&Stars 1 and 2 merge, 3 remains bound\cr
&\tt merger\_escape\_1&Stars 2 and 3 merge, 1 escapes\cr
&\tt merger\_escape\_2&Stars 1 and 3 merge, 2 escapes\cr
&\tt merger\_escape\_3&Stars 1 and 2 merge, 3 escapes\cr
&\tt triple\_merger&All three stars merge\cr
\noalign{\smallskip\hrule}
}}
\smallskip\noindent
$^\dagger$The state names are those used by the STARLAB scattering
software.

\par\vfill\eject

\null\bigskip\noindent
{\bf Figure Captions}

\medskip\noindent
{\bf Figure 1:}~~Scaled cross sections for ``clean'' (i.e.
non-colliding) exchange (open circles), ionization (stars), two-star
mergers (filled circles), and three-star mergers (filled triangles),
as functions of initial binary semi-major axis for the particular
binary--single-star interaction described in the text.  Error bars
represent formal 1-$\sigma$ errors, and are shown only where they
exceed the size of the symbol used to represent the data.

\medskip\noindent
{\bf Figure 2:}~~(a) Branching ratios within the class of ``clean
exchanges'' (Fig.~1) for ejection of star 1 (squares) and star 2
(hexagons).  (b) Further subdivision of the exchange cross section to
indicate resonance (grey filled) and non-resonance (open) encounters.

\medskip\noindent
{\bf Figure 3:}~~Branching ratios within the class of all mergers
(Fig.~1) for mergers of stars 1 and 2 (the initial binary components;
squares), stars 1 and 3 (pentagons) and stars 2 and 3 (hexagons).
Also shown (triangles) is the triple merger cross section, normalized
to the total 2-merger cross section.

\medskip\noindent
{\bf Figure 4:}~~Branching ratios within the class of two-body mergers
(Fig.~1) for the formation of unbound merger products.  For small $a$,
many mergers occur in resonant encounters, leading to a system in
which the merger product is bound to the third star in the (initial)
system.  For large $a$, mergers are overwhelmingly of the ``induced''
type, resulting in unbound merger products.  As in Fig.~3, mergers of
stars 1 and 2 are shown as squares, 1 and 3 as pentagons, 2 and 3 as
hexagons.  The total unbound merger probability is indicated by
circles.

\bye